\def\fracm#1#2{\hbox{\large{${\frac{{#1}}{{#2}}}$}}}

\documentstyle[12pt]{article}

\skewchar\fivmi='177
\skewchar\sixmi='177
\skewchar\sevmi='177
\skewchar\egtmi='177
\skewchar\ninmi='177
\skewchar\tenmi='177
\skewchar\elvmi='177
\skewchar\twlmi='177
\skewchar\frtnmi='177
\skewchar\svtnmi='177
\skewchar\twtymi='177
\def\@magscale#1{ scaled \magstep #1}
\skewchar\fivsy='60
\skewchar\sixsy='60
\skewchar\sevsy='60
\skewchar\egtsy='60
\skewchar\ninsy='60
\skewchar\tensy='60
\skewchar\elvsy='60
\skewchar\twlsy='60
\skewchar\frtnsy='60
\skewchar\svtnsy='60
\skewchar\twtysy='60

\catcode`@=11
\def\un#1{\relax\ifmmode\@@underline#1\else
        $\@@underline{\hbox{#1}}$\relax\fi}
\catcode`@=12

\def\a{\alpha}
\def\b{\beta}

\def\d{\delta}
\def\e{\epsilon}

\def\g{\gamma}

\def\l{\lambda}

\def\s{\sigma}
\def\t{\tau}

\def\dslash{\not{\hbox{\kern-2pt $\partial$}}}
\def\Dslash{\not{\hbox{\kern-4pt $D$}}}
\def\pslash{\not{\hbox{\kern-2.3pt $p$}}}
 \newtoks\slashfraction
 \slashfraction={.13}
 \def\slash#1{\setbox0\hbox{$ #1 $}
 \setbox0\hbox to \the\slashfraction\wd0{\hss \box0}/\box0 }

\font\ro=cmsy10                          
\def\kcr{{\hbox{\ro \char'170}}}                
\def\ktl{{\hbox{\ro \char'170}}}        
\def\ktr{{\hbox{\ro \char'170}}}        
\def\kbl{{\hbox{\ro \char'170}}}        
\def\kbr{{\hbox{\ro \char'170}}}        

\def\plpl{\raise-2pt\hbox{$\raise3pt\hbox{$_+$}\hskip-6.67pt\raise0.0pt
\hbox{$^+$}\hskip 0.01pt$}}
\def\mimi{\raise-2pt\hbox{$\raise3pt\hbox{$_-$}\hskip-6.67pt\raise0.0pt
\hbox{$^-$}\hskip 0.01pt$}}

\def\bo{{\raise.15ex\hbox{\large$\Box$}}}               
\def\pa{\partial}                                       
\def\TH{{\raise.2ex\hbox{$\displaystyle \bigodot$}\mskip-4.7mu \llap H \;}}
\def\face{{\raise.2ex\hbox{$\displaystyle \bigodot$}\mskip-2.2mu \llap {$\ddot
        \smile$}}}                                      

\def\sp#1{{}^{#1}}                              
\def\leftrightarrowfill{$\mathsurround=0pt \mathord\leftarrow \mkern-6mu
        \cleaders\hbox{$\mkern-2mu \mathord- \mkern-2mu$}\hfill
        \mkern-6mu \mathord\rightarrow$}
\def\dvec#1{\vbox{\ialign{##\crcr
        \leftrightarrowfill\crcr\noalign{\kern-1pt\nointerlineskip}
        $\hfil\displaystyle{#1}\hfil$\crcr}}}           

\def\fracm#1#2{\hbox{\large{${\frac{{#1}}{{#2}}}$}}}
\def\frac#1#2{{\textstyle{#1\over\vphantom2\smash{\raise.20ex
        \hbox{$\scriptstyle{#2}$}}}}}                   
\def\sfrac#1#2{{\vphantom1\smash{\lower.5ex\hbox{\small$#1$}}\over
        \vphantom1\smash{\raise.4ex\hbox{\small$#2$}}}} 
\def\bfrac#1#2{{\vphantom1\smash{\lower.5ex\hbox{$#1$}}\over
        \vphantom1\smash{\raise.3ex\hbox{$#2$}}}}       
\def\afrac#1#2{{\vphantom1\smash{\lower.5ex\hbox{$#1$}}\over#2}}    

\newskip\humongous \humongous=0pt plus 1000pt minus 1000pt
\def\caja{\mathsurround=0pt}
\def\eqalign#1{\,\vcenter{\openup2\jot \caja
        \ialign{\strut \hfil$\displaystyle{##}$&$
        \displaystyle{{}##}$\hfil\crcr#1\crcr}}\,}
\newif\ifdtup

\def\ref#1{$\sp{#1)}$}

\parskip=\medskipamount                 
\thicklines                         

\def\oldheadpic{                                
        \setlength{\unitlength}{.4mm}
        \thinlines
        \par
        \begin{picture}(349,16)
        \put(325,16){\line(1,0){4}}
        \put(330,16){\line(1,0){4}}
        \put(340,16){\line(1,0){4}}
        \put(335,0){\line(1,0){4}}
        \put(340,0){\line(1,0){4}}
        \put(345,0){\line(1,0){4}}
        \put(329,0){\line(0,1){16}}
        \put(330,0){\line(0,1){16}}
        \put(339,0){\line(0,1){16}}
        \put(340,0){\line(0,1){16}}
        \put(344,0){\line(0,1){16}}
        \put(345,0){\line(0,1){16}}
        \put(329,16){\oval(8,32)[bl]}
        \put(330,16){\oval(8,32)[br]}
        \put(339,0){\oval(8,32)[tl]}
        \put(345,0){\oval(8,32)[tr]}
        \end{picture}
        \par
        \thicklines
        \vskip.2in}
\def\oldtitle#1#2#3#4{\oldheadpic\begin{center}\vglue.5in{\large\bf #1}\\[.6in]
        {#2}\\[.1in] {\it Department of Physics and Astronomy}\\
        {\it University of Maryland, College Park, MD 20742}\\[.6in]
        Physics Publication \#{#3}\\ {#4}\\[1.5in] {\bf ABSTRACT}\\[.1in]
        \end{center} \begin{quotation}}                 
\def\oldTitle#1#2#3#4#5#6#7{\oldheadpic\begin{center} \vglue .4in
        {\large\bf #1}\\[.4in]
        {#2}\\[.1in] {\it Department of Physics and Astronomy}\\
        {\it University of Maryland, College Park, MD 20742}\\[.1in]
        {#3}\\[.1in] {\it {#4}}\\ {\it {#5}}\\[.4in]
        Physics Publication \#{#6}\\ {#7}\\[.5in] {\bf ABSTRACT}\\[.1in]
        \end{center} \begin{quotation}}                 
\def\border{                                            
        \setlength{\unitlength}{1mm}
        \newcount\xco
        \newcount\yco
        \xco=-21
        \yco=12
        \begin{picture}(140,0)
        \put(\xco,\yco){$\ktl$}
        \advance\yco by-1
        {\loop
        \put(\xco,\yco){$\kcr$}
        \advance\yco by-2
        \ifnum\yco>-240
        \repeat
        \put(\xco,\yco){$\kbl$}}
        \xco=158
        \yco=12
        \put(\xco,\yco){$\ktr$}
        \advance\yco by-1
        {\loop
        \put(\xco,\yco){$\kcr$}
        \advance\yco by-2
        \ifnum\yco>-240
        \repeat
        \put(\xco,\yco){$\kbr$}}
        \put(-20,13){\tiny University of Maryland Elementary Particle
Physics University of Maryland Elementary Particle Physics University of
Maryland Elementary Particle Physics}
        \put(-20,-241.5){\tiny University of Maryland Elementary
Particle Physics University of Maryland Elementary Particle Physics
University of Maryland Elementary Particle Physics}
        \end{picture}
        \par\vskip-8mm}
\def\bordero{                                           
        \setlength{\unitlength}{1mm}
        \newcount\xco
        \newcount\yco
        \xco=-31
        \yco=12
        \begin{picture}(140,0)
        \put(\xco,\yco){$\ktl$}
        \advance\yco by-1
        {\loop
        \put(\xco,\yco){$\kclr}
        \advance\yco by-2
        \ifnum\yco>-240
        \repeat
        \put(\xco,\yco){$\kbl$}}
        \xco=151
        \yco=12
        \put(\xco,\yco){$\ktr$}
        \advance\yco by-1
        {\loop
        \put(\xco,\yco){$\kcr$}
        \advance\yco by-2
        \ifnum\yco>-240
        \repeat
        \put(\xco,\yco){$\kbr$}}
        \put(-20,12){\ooo
bacdefghidfghghdhededbihdgdfdfhhdheidhdhebaaahjhhdahba

hgdedge
   hgfdiehhgdigicba}
        \put(-20,-241.5){\ooo
ababaighefdbfghgeahgdfgafagihdidihiidhiagfedhadbfd

ecdcdfa
   gdcbhaddhbgfchbgfdacfediacbabab}
        \end{picture}
        \par\vskip-8mm}
\def\headpic{                                           
        \indent
        \setlength{\unitlength}{.4mm}
        \thinlines
        \par
        \begin{picture}(29,16)
        \put(165,16){\line(1,0){4}}
        \put(170,16){\line(1,0){4}}
        \put(180,16){\line(1,0){4}}
        \put(175,0){\line(1,0){4}}
        \put(180,0){\line(1,0){4}}
        \put(185,0){\line(1,0){4}}
        \put(169,0){\line(0,1){16}}
        \put(170,0){\line(0,1){16}}
        \put(179,0){\line(0,1){16}}
        \put(180,0){\line(0,1){16}}
        \put(184,0){\line(0,1){16}}
        \put(185,0){\line(0,1){16}}
        \put(169,16){\oval(8,32)[bl]}
        \put(170,16){\oval(8,32)[br]}
        \put(179,0){\oval(8,32)[tl]}
        \put(185,0){\oval(8,32)[tr]}
        \end{picture}
        \par\vskip-6.5mm
        \thicklines}
\def\title#1#2#3#4{\border\headpic {\hbox to\hsize{#4 \hfill UMDEPP #3}}\par
        \begin{center} \vglue .5in {\large\bf #1}\\[.6in]
        {#2}\\[.1in] {\it Department of Physics and Astronomy}\\
        {\it University of Maryland, College Park, MD 20742}\\[1.5in]
        {\bf ABSTRACT}\\[.1in] \end{center} \begin{quotation}}  
\def\Title#1#2#3#4#5#6#7{\border\headpic
        {\hbox to\hsize{#7 \hfill UMDEPP #6}}\par
        \begin{center} \vglue .4in {\large\bf #1}\\[.4in]
        {#2}\\[.1in] {\it Department of Physics and Astronomy}\\
        {\it University of Maryland, College Park, MD 20742}\\[.1in]
        {#3}\\[.1in] {\it {#4}}\\ {\it {#5}}\\[.5in] {\bf ABSTRACT}\\[.1in]
        \end{center} \begin{quotation}}                 
\def\endtitle{\end{quotation}\newpage}                  

\def\sect#1{\bigskip\medskip \goodbreak \noindent{\bf {#1}} \nobreak \medskip}

\begin{document}

\def\gfrac#1#2{\frac {\scriptstyle{#1}}
        {\mbox{\raisebox{-.6ex}{$\scriptstyle{#2}$}}}}
\def\gg{{\hbox{\sc g}}}
\border\headpic {\hbox to\hsize{March 1995 \hfill {UMDEPP 95-113}}}
\par
\setlength{\oddsidemargin}{0.3in}
\setlength{\evensidemargin}{-0.3in}
\begin{center}
\vglue .08in
{\large\bf A Theory of Spinning Particles for Large\\
N-extended Supersymmetry}
\\[.72in]

S. James Gates, Jr.\footnote {Supported in part by National Science
Foundation Grant
PHY-91-19746}
${}^,$ \footnote {Supported in part by NATO Grant CRG-93-0789}
and L. Rana
\\[.02in]
{\it Department of Physics\\
University of Maryland\\
College Park, MD 20742-4111  USA}\\
{\bf {\tt gates@umdhep.umd.edu}}\\
{\bf {\tt lubna@umdhep.umd.edu}}\\[1.2in]

{\bf ABSTRACT}\\[.002in]
\end{center}
\begin{quotation}
{An explicit construction of theories of spinning particles, both
massive and massless, is given with arbitrary extended supersymmetry
on the world-line. As an application of our results, we give a
{\underline {universal}} description of 3D (and via truncation
all lower dimensional) supersymmetric scalar multiplets with
arbitrary N-extended supersymmetry. }

\endtitle
\section{Introduction}

{}~~~~The simplest of supersymmetric systems are those that involve
a single bosonic dimension. The best known example of this is the
spinning particle \cite{SPART}. The uses of such systems is
surprisingly broad given their simplicity.  Some of these include
simulation of more complicated spinning strings \cite{EX1}, calculation
of field theory anomalies \cite{EX2}, possible realization of new
Kac-Moody algebras \cite{EX3} and as generating functions for
superintegrable systems like SKdV \cite{EX4}.  Despite all of these
applications, a search of the literature reveals that almost no
work has been done in elucidating the general structure of such
theories.  This will be the purpose of this work.

In the following we will give an explicit construction of the
spinning particle action (both massive and massless) for all values
of N-extended supersymmetry. We show that a basis for all
such theories is provided by certain real representations of
generalized Pauli-matrix algebras.   As an application of our work,
we construct a universal description of on-shell scalar
multiplets with arbitrary N-extended supersymmetry in three or less
bosonic dimensions.

\section{Minimal 1D, N-Extended Supergravity}

{}~~~~In this section, we introduce the concept of a minimal 1D
representation of the supergravity multiplet that is valid for
arbitrary (and large) values of N, the degree of extended supersymmetry.
The  supersymmetry variations of the 1D, N-extended worldline
supergravity fields may be defined on
($e \, , \, \chi^{\rm I}$) where $e(\t)$ is real and $\chi^{\rm I}(\t)$
is a real spinor (with ${\rm I} = 1, ..., {\rm N}$),
$$
\d_{Q} e ~=~ - i \, 4 e^{2}  \, \a^{\rm I} \, {\chi}^{\rm I} \, ~~~, ~~~
\d_{Q} \chi^{\rm I}~=~ - \left( \pa_{\t} \a^{\rm I} \right)  \, ~~~.
\eqno(2.1)$$
The general coordinate transformations can be written as:
$$
\d_{G.C} e~=~\left( \pa_{\t} e  \right) \xi ~-~ e \, \left( \pa_{\t} \xi
\right) ~~~, ~~~
\d_{G.C} \chi^{\rm I}~=~ \pa_{\t} (\,  \chi^{\rm I} \xi \,) ~~~.
 \eqno(2.2) $$
Now a direct calculation shows that
$$ \eqalign{
\left[ \d_{Q} \left( \a_{1} \right), \d_{Q} \left( \a_{2} \right)
\right]~~~&=~~~\d_{G.C.}(\xi_{12}) ~+~ \d_{Q}({\a_{12}}^{\rm I})
{}~~~.  \cr } \eqno(2.3) $$
where ${\xi}_{12} \equiv i 4 e (\, \a_1 {}^{\rm I} \a_2 {}^{\rm I} \,) $
and ${\a_{12}}^{\rm I} \equiv \xi_{12} \chi^{\rm I} $.
Thus, it is not necessary to gauge a set of automorphisms (i.e. the
generalization of the U(1) group seen in  the N = 2 nonminimal theory)
and (2.3) may be taken as the definition of  the 1D local supersymmetry
group for arbitrary numbers of real supercharges.

\section{1D, N-Extended Rigid Supersymmetry and \newline Matter}

{}~~~~Having begun with the representation of the supergravity multiplet
for arbitrary values of N, we next need to study the matter multiplets
that are available to be coupled to these minimal worldline supergravity
multiplets.  In terms of real fields ($\phi_{i} , \, \psi_{
\hat j}$) with $i = 1,...,{\rm d}$, ${\hat i} = 1,...,{\rm d}$
we propose that the supersymmetry transformation laws for a scalar
supermultiplet can be written as:
$$
\d_{Q} \, \phi_{i}~=~i \, 2 \a^{K} \, \left( {\rm L}_{K} \right)_i
{}^{\hat j} \,  \psi_{\hat j} ~~~, ~~~
\d_{Q} \, \psi_{\hat i}~=~ \a^{K} \, \left( {\rm R}_{K} \right)_{\hat i}
{}^{j} \,
\pa_{\t} \phi_{j} ~~~ .  \eqno(3.1) $$
It is of importance that we have not made the {\it a} {\it {priori}} assumption
that ${\rm d} = {\rm N}$.  Note that the reality of the fields of necessity
implies that the matrices $\left( {\rm L}_{K} \right)_i {}^{\hat j}$ and $
\left( {\rm R}_{K} \right)_{\hat i} {}^{j}$ that appear above are also real.
By requiring that the operator equation should read
$$\eqalign{
\left[ \d_{Q} \left( \a_{1} \right), \d_{Q}\left( \a_{2}
\right)
 \right]~~~&=~~  \,i 4 \, \a_1 {}^{K} \, \a_2 {}^{K} \pa_{\t} ~~~ , }
\eqno(3.2)$$
uniformly on both fields we find that the matrices L and R satisfy the
following
algebra:
$$ {\rm L}_{K} \, {\rm R}_{P} \, + \, {\rm L}_{P} \, {\rm R}_{K}~~=~~- \, 2 \,
\d_{KP} \,
{\rm I} ~~~, ~~~
{\rm R}_{K} \, {\rm L}_{P} \, + \, {\rm R}_{P} \, {\rm L}_{K}~~=~~- \, 2 \,
\d_{KP} \,
{\rm I} ~~~.
\eqno(3.3) $$

If the representations denoted by $i$ are distinguishable from those
denoted by ${\hat i}$, it follows that a second distinct ``twisted'' scalar
multiplet can be constructed. It is described by the set of supersymmetry
variations given by
$$
\d_{Q} \, {\widehat \phi}_{\hat i}~=~ i \, 2 \a^{K} \, \left( {\rm R}_{K}
\right)_{\hat i} {}^{j} \, {\widehat \psi}_{ j} ~~~,~~~
\d_{Q} \, {\widehat \psi}_{i}~=~ \a^{K} \, \left( {\rm L}_{K} \right)_{i}
{}^{\hat j} \, \pa_{\t} {\widehat \phi}_{\hat j} ~~~.  \eqno(3.4) $$
Furthermore, it may happen that there are a number of {\underline
{inequivalent}}
representations of the algebra in (3.3). For each such inequivalent
representation,
there would correspond a distinct scalar multiplet such as appears in (3.1)
together with a twisted version such as in (3.4)\footnote{This is the origin
of the surprisingly large number of (4,0) scalar and spinor multiplets
found in \newline ${~~~~~}$ \cite{GaLu}.}.

For general values of N, we can also construct spinor multiplets with the
respective transformations laws,

$$
\d_{Q} \, {\eta}_{\hat i}~=~  \a^{K} \, \left( {\rm R}_{K}
\right)_{\hat i} {}^{j} \, { F}_{ j} ~~~, ~~~
\d_{Q} \, { F}_{i} ~=~ i 2 \, \a^{K} \, \left( {\rm L}_{K} \right)_{i}
{}^{\hat j} \, \pa_{\t} { \eta}_{\hat j} ~~~.  \eqno(3.5) $$

and twisted spinor multiplets
$$
\d_{Q} \, { \widehat \eta}_{i}~=~ \a^{K} \,
 \left( {\rm L}_{K} \right)_i {}^{\hat j}
 \, {\widehat F}_{\hat j} ~~~, ~~~
\d_{Q} \, {\widehat F}_{\hat i}~=~ i2 \, \a^{K} \,
\left( {\rm R}_{K} \right)_{\hat i} {}^{j} \,
\pa_{\t} {\widehat \eta}_{j} ~~~. \eqno(3.6) $$

Again the number of such distinct multiplets is in one-to-one correspondence
with the number of inequivalent representation of the algebra in (3.3).

Before we move on to the question of invariant actions, it is extremely
interesting to note that the existence of the 1D, arbitrary N-extended
supergravity multiplet was {\underline {independent}} of the existence of
the algebra in (3.3).  The 1D supergravity formed a representation of the
local 1D supersymmetry algebra in (2.3) without reference to the matrix
algebra in (3.3). It is the existence of the matter field representations
that depend on the latter algebra.  We believe that this is an important
lesson regarding representations of local supersymmetry in general.  The
local 1D supergravity multiplet need not be tied to any matrix realization of
supersymmetry\footnote{It is tempting to believe that the present inability
to find off-shell representations of 4D, $N > 4$ \newline ${~~~~~}$
supergravity
is somehow related to this fact.}!  On the other hand, the matter multiplets
can be classified by precisely the number of inequivalent representations of
the matrix representation.

For an invariant Lagrangian, we may take
$$
{\cal L}_{SM} ~=~   (\, {\pa}_{\t} \phi_{i} \,) \,
(\, {\pa}_{\t} \phi_{i} \,) \, + \, i 2 \psi_{\hat i} {\pa}_{\t} \psi_{\hat i}
 ~~~, \eqno(3.7)
$$
whose invariance requires the condition,
$$ ({\rm L}_{I})^{i \hat j} ~+~ ({\rm R}_{I})^{\hat j i} ~=~ 0 ~~~,
\eqno(3.8)
$$
where the indices on the ${\rm L}$ and ${\rm R}$ matrices are raised with
the use of Kronecker deltas.

In a similar manner, we can write an invariant Lagrangian for the spinor
multiplet of the form,
$${\cal L}_{FM} ~=~ i 2 \, \eta_{\hat i} \, {\pa}_{\t} \, \eta_{\hat i}
{}~+~ F_{i} \, F_{i}  ~~~,  \eqno(3.9)$$
whose invariance leads to precisely the same condition as in (3.8).  The
same identity arises if one considers the twisted versions of the multiplets.

\section{1D, N-Extended Local Systems}

{}~~~~Having determined the general structure of 1D supergravity and as well
the
general structure of 1D supersymmetric scalar and spinor multiplets, it is now
time to put these two pieces together and investigate spinning particle
theories. To begin the local version of (3.1) takes the form
$$
\d_{Q} \, \phi_{i}~=~i 2 \, \a^{K} \, \left( {\rm L}_{K} \right)_i {}^{\hat j}
 \, \psi_{\hat j} ~~~, ~~~
\d_{Q} \, \psi_{\hat i}~=~ \a^{K} \, \left( {\rm R}_{K} \right)_{\hat i} {}^{j}
 \, {\cal D}_{\t} \phi_{j} ~~~,  \eqno(4.1) $$
and leave invariant the action;
$${\cal S}_{SM} = \int d {\t} \, [~ e^{-1} \,{\cal D}_{\t} \phi_{i} \,
{\cal D}_{\t} \phi_{i} \, + \, i 2 \, e^{-1} \psi_{\hat i} {\cal D}_{\t}
\psi_{\hat i} \, + \, i 2 \, \chi^{K} \left( {\rm R}_{K} \right)_{\hat i} {}^j
\, \psi_{\hat i} \, {\cal D}_{\t} \phi_{j} ~] ~~~, \eqno(4.2)$$
where
$$ \eqalign {
{\cal D}_{\t} \phi_{i} &\equiv ~ e \, [~ \pa_{\t} \phi_{i} ~+~ i 2 \chi^{K}
\left( {\rm L}_{K} \right)_i {}^{\hat k} \psi_{\hat k} ~] ~~~,  \cr
{\cal D}_{\t} \psi_{\hat k} &\equiv ~ e \,[~ \pa_{\t} \psi_{\hat k} ~+~
\chi^{K} \left( {\rm R}_{K} \right)_{\hat k} {}^{i} {\cal D}_{\t} \phi_i ~]
{}~~~.  }  \eqno(4.3) $$

The local transformation laws for the spinor multiplet are given by
$$
\d_{Q} \, \eta_{\hat i}~=~  \a^{K} \, \left( {\rm R}_{K} \right)_{\hat i}
{}^{j}
\, F_{j} ~~~, ~~~
\d_{Q} \, F_{i}~=~ i2\a^{K} \, \left( {\rm L}_{K} \right)_i {}^{\hat j} \,
{\cal D}_{\t} \eta_{\hat j}~~~,  \eqno(4.4) $$
where
$$ {\cal D}_{\t} \eta_{\hat i} \equiv ~ e \, [~ \pa_{\t} \eta_{\hat i} ~+~
 \chi^{K}
 \left( {\rm R}_{K} \right)_{\hat i} {}^{k} F_{k} ~] ~~~. \eqno(4.5)
$$
An invariant action is given by
$${\cal S}_{FM} = \int d {\t} \, [~ e^{-1} \, i 2 \, \eta_{\hat i}
{\cal D}_{\t} \eta_{\hat i} ~ + ~ e^{-1} F_{i}  \, F_{i}  ~ + ~
i 2  \, \chi^{K} \left( {\rm R}_{K} \right)_{\hat i} {}^j \,
\eta_{\hat i} \, F_{j}  ~] ~~~,  \eqno(4.6)$$
whose relevance will become clear below.

It is, of course, possible to consider these supermultiplets as representations
of some additional symmetry groups.  Under this circumstance each field of the
multiplet carries an additional index $ ( \eta_{\hat i} \, , \, F_{i} ) \to
(\eta_{\hat i} {}^{\hat {\rm I}} \, , \, F_{i} {}^{\hat {\rm I}} )$.
In the construction
below, this property will be important. In particular, we can pick ${\hat {\rm
I}} = i$, so that the auxiliary field becomes a two by two matrix $
F_i {}^j$ which can appear in the following invariant actions;
$${\widetilde {\cal S}}_{FM} ~=~ \int d {\t} \, [~ e^{-1} \,
 \, i 2 \, \eta_{\hat i} {}^{k} {\cal D}_{\t} \eta_{\hat i} {}^{k}
\, + \, e^{-1} F_{i} {}^{k} \, F_{i} {}^{k}  \, + \,
i 2  \, \chi^{K} \left( {\rm R}_{K} \right)_{\hat i} {}^{j} \,
\eta_{\hat i} {}^{k} \, F_{j} {}^{k} ~] ~~~,  \eqno(4.7) $$
and
$${\cal S}_{Cosm} ~=~ m_0 \int d {\t} \, [~ e^{-1} \, F_{i} {}^i  \, - \, i
2 \, \chi^{K} \left( {\rm L}_{K} \right)_{i} {}^{\hat j} \, \eta_{\hat j} {}^i
{}~] ~~~,  \eqno(4.8)$$

Finally, there is a term describing an interaction between a scalar
multiplet and a single spinor multiplet
$${\cal S}_{Int} = \int d {\t} \, [~ e^{-1} \,  F_i \,
{\cal D}_{\t} \phi_{i} \, + \, i 2 e^{-1} \eta_{\hat i} \,
{\cal D}_{\t} \psi_{\hat i} \, + \,
i 2  \, \chi^{K} \left( {\rm R}_{K} \right)_{\hat i} {}^{j} \,
\eta_{\hat i}  \, {\cal D}_{\t} \phi_j
 ~] ~~~.  \eqno(4.9)$$

We first consider the case of the massless spinning particle with
arbitrary N-extended supersymmetry in a second order formalism. The action
for such a theory is precisely ${\cal S}_{SM}$. The scalar field $\phi_i
(\t)$ represents the coordinate of the particle in a ${\rm d}$-dimensional
spacetime. The quantity $\psi_{\hat i} (\t)$ is the NSR fermion. The first
order
formalism is obtained by taking as an action ${\cal S}_{Int} \, - \, \fracm 12
{\cal S}_{FM}$. One of the interesting points of the first order formalism
is that the auxiliary field of the spinor multiplet, $F_i (\t)$, plays
the role of the momentum canonically conjugate to the coordinate $\phi_i
(\t)$. Thus, the entire spinor multiplet ($\eta_{\hat i} ,~  F_i $) is the
canonical conjugate to the scalar multiplet ($\phi_i , ~ \psi_{\hat i} $).

On the other hand, the massive versions of the theory are considerably
different.  To describe this version requires that we actually add to
either the first or second order massless actions an additional spinor
multiplet action!  This mass term is describe by ${\rm d}^{-1} [ \,
{\cal S}_{Cosm} - \fracm 12 {\widetilde {\cal S}}_{FM} \, ]$.  The
equation of motion for $F_i {}^i$ sets it equal to the mass parameter
$ m_0$.  Without the extra spinor multiplet ($\eta_{\hat k} {}^j ,~  F_i
{}^j $), it is not possible to introduce the mass. Finally note that the
twisted version of all of the above actions can also be easily constructed.

\section{Conclusions and Summary}

{}~~~~We have shown that it is possible to construct spinning particles
with arbitrary numbers of supersymmetries on the world-line. This result
has a number of interesting implications.  For example, the N = 1 and N
= 2 SKdV systems have been shown to arise precisely by starting from
the massive spinning particle action and performing a field redefinition on
the equation of motion of the scalar multiplet to derive a supersymmetric
Lax equation. The implication of our work is that it is possible to infinitely
extend the amount of supersymmetry permitted as generalizations of the KdV
system!

Another course that is indicated as a direction of further
investigation is the study of utilizing 1D locally supersymmetric systems
in descriptions of proposed hyperbolic  Kac-Moody algebras as first
envisioned by Julia.  Here our results for N = 16 may be taken as
a point of departure. The topic of locally supersymmetric non-linear
1D sigma-models is clearly one that can now be studied in a much more
systematic manner.  Finally, the interesting question arises as to
whether it is possible to take the N $\to \infty$ limit of these
1D models? If this is possible, do we find a new super Kac-Moody
algebra?

\sect{APPENDIX A: Explicit {\rm L} and {\rm R} Matrix Representations}

In the text of this paper, we have seen that the problem of constructing
the 1D, N-extended spinning particle action reduces to finding real matrix
solutions to the equations of (3.3) and (3.8).

A more complete discussion of the existence of solutions for these conditions
in general will be included in a forthcoming work \cite{LuGA}.  Here we
provide some examples of solutions to these conditions for values up to
N = 16.

\noindent For N = 2, there exist a set of two by two matrices:
$$\eqalign{
{\rm L}_{1} ~=~ i \s^2 ~=~ {\rm R}_{1}  ~~~ ; ~~~
{\rm L}_{2} ~=~ - \, {\rm I} ~=~ - \, {\rm R}_{2} ~~~~ . } \eqno(A.1) $$

\noindent For N = 4, there are two distinct minimal set of matrices that
realize
the algebra and are well known in the physics literature
$$ \begin{array}{ccccccccccccccc}
{\rm L}_{1} &=& i \s^1 & \otimes & \s^2 &=& {\rm R}_{1} & ~~~; ~~~ &
{\widehat {\rm L}}_1 &=& i \s^2 & \otimes  & \s^3 &=&
{\widehat {\rm R}}_1 ~~~ ;  \\
{\rm L}_{2} &=& i \s^2 & \otimes & {\rm I} &=& {\rm R}_{2} & ~~~ ; ~~~ &
{\widehat{\rm L}}_2 &=& - i {\rm I} & \otimes  & \s^2 &=&
{\widehat {\rm R}}_2 ~~~ ;  \\
{\rm L}_{3} &=& -i \s^3 & \otimes & \s^2 &=& {\rm R}_{3} & ~~~ ; ~~~ &
{\widehat {\rm L}}_3 &=& i \s^2  & \otimes & \s^1 &=&
{\widehat {\rm R}}_3 ~~~ ;  \\
{\rm L}_{4} &=&  \, {\rm I} & \otimes & {\rm I} &=& - \, {\rm R}_{4} & ~~~;~~~
&
{\widehat {\rm L}}_{4} &=&  \, {\rm I} & \otimes & {\rm I} &=& - \,
{\widehat {\rm R}}_{4} ~~~.
\end{array} \eqno(A.2) $$
These provide an example of a set of inequivalent representations. Any three
within a given set can be used to cover the case of N = 3.

\noindent For N = 8, a convenient set for our required matrices is given by,
$$  \begin{array}{ccccccccccccccccccc}
{\rm L}_{1} &=& i {\rm I} & \otimes & \s^3 & \otimes & \s^2 &=& {\rm R}_{1}
&~~~; ~~~& {\rm L}_{5} &=& i \s^2 & \otimes & {\rm I}  & \otimes &  \s^1
&=& {\rm R}_{5} ~~~~; \\
{\rm L}_{2} &=& i\s^3 & \otimes & \s^2 & \otimes & {\rm I} &=& {\rm R}_{2}
&~~~; ~~~& {\rm L}_{6} &=& i \s^2 & \otimes & {\rm I} & \otimes & \s^3
&=& {\rm R}_{6} ~~~~; \\
{\rm L}_{3} &=& i {\rm I} & \otimes & \s^1 & \otimes & \s^2 &=& {\rm R}_{3}
&~~~; ~~~ & {\rm L}_{7} &=& i \s^2 & \otimes & \s^2 & \otimes & \s^2
&=& {\rm R}_{7}  ~~~~; \\
{\rm L}_{4} &=& i \s^1 & \otimes & \s^2 & \otimes & {\rm I} &=& {\rm R}_{4}
&~~~; ~~~ & {\rm L}_{8} &=& \,  {\rm I}\ & \otimes & {\rm I} & \otimes & {\rm
I}
&=& - \, {\rm R}_{8} ~  .
\end{array}
 \eqno(A.3) $$
N = 5, 6, and 7 can be formed by taking any set of 5, 6 or 7 of the N = 8
matrices,
respectively.

\noindent For N = 9, we find the matrices to be 16 x 16:
$$ \begin{array}{ccccccccccc}
 L_1&=& i {\rm I} & \otimes & \s^3 & \otimes & \s^2 & \otimes & \s^1 & =& R_1
\\
L_2&=& i \s^3 & \otimes & \s^2 & \otimes &{\rm I} & \otimes & {\rm I}& =& R_2
\\
L_3 &=& i{\rm I} & \otimes & \s^1 & \otimes &  \s^2 & \otimes & \s^1& =& R_3
\\
L_4&=& i \s^1 & \otimes & \s^2 & \otimes & {\rm I} & \otimes &{\rm I} &=& R_4
\\
L_5&=& i \s^2 & \otimes & {\rm I} & \otimes &  \s^1 & \otimes & \s^1& =& R_5
\\
L_6&=& i \s^2 & \otimes & {\rm I} & \otimes & \s^3 & \otimes & \s^1& =& R_6
\\
L_7&=& i \s^2 & \otimes & \s^2 & \otimes & \s^2 & \otimes & \s^1& =& R_7
\\
L_8&=& i \s^2 & \otimes & {\rm I} & \otimes & {\rm I} & \otimes & \s^3& =& R_8
\\
L_9&=& {\rm I}& \otimes & {\rm I} & \otimes & {\rm I}&\otimes & {\rm I}&=&
- R_9
\end{array} \eqno(A.4) $$

\noindent For N =10, we have been able to find a 32 x 32 representation:
$$\begin{array}{cccccccccccccrc}
{\rm L}_{1} &=& i \s^2 &\otimes & {\rm I} & \otimes & {\rm I}& \otimes &
{\rm I} & \otimes &{\rm I} & =&{\rm R}_{1}& ; \\
{\rm L}_{2} &=& i \s^1 &\otimes &\s^2 &\otimes& {\rm I}& \otimes &{\rm I}&
\otimes & {\rm I}& =&{\rm R}_{2}&; \\
{\rm L}_{3}  &=& i \s^1 &\otimes &\s^1 &\otimes &\s^2 &\otimes & {\rm I}&
\otimes & {\rm I} &=&{\rm R}_{3}&;  \\
{\rm L}_{4} &=& i \s^1& \otimes &\s^3 &\otimes &\s^2 &\otimes & {\rm I}&
\otimes
& {\rm I} &=&{\rm R}_{4}&;  \\
{\rm L}_{5} &=& i \s^3 &\otimes & {\rm I} & \otimes & \s^2 & \otimes &
\s^1 & \otimes &{\rm I} & =&{\rm R}_{5}& ; \\
{\rm L}_{6} &=& i \s^3 &\otimes & {\rm I} & \otimes & \s^2& \otimes &
\s^3 & \otimes &{\rm I} & =&{\rm R}_{6}& ; \\
{\rm L}_{7} &=& i \s^3&\otimes & {\rm I} & \otimes & {\rm I}& \otimes &
\s^2 & \otimes &\s^1 & =&{\rm R}_{7}& ; \\
{\rm L}_{1} &=& i \s^3 &\otimes & {\rm I} & \otimes & \s^2 & \otimes &
\s^2 & \otimes &\s^2 & =&{\rm R}_{8}& ; \\
{\rm L}_{9} &=& i \s^3 &\otimes & {\rm I} & \otimes & {\rm I}& \otimes &
\s^2 & \otimes &\s^3 & =&{\rm R}_{9}& ; \\
{\rm L}_{10} &=&{\rm I} &\otimes & {\rm I} & \otimes & {\rm I}& \otimes &
{\rm I} & \otimes &{\rm I} & =&-{\rm R}_{10}& ;
 \end{array} \eqno(A.5) $$

\noindent N = 12 is a 64 x 64 representation:
$$
\begin{array}{ccccccccccccccc}
L_1& = &i \s^2 & \otimes &{\rm I}& \otimes & {\rm I} & \otimes &{\rm I} &
\otimes &
{\rm I} & \otimes & {\rm I} & = & {\rm R}_1 ~~~~~~~ ; \\
{\rm L}_2 & = & i \s^1 & \otimes & \s^2 & \otimes & {\rm I} & \otimes & {\rm I}
& \otimes
& {\rm I} & \otimes & {\rm I} & = & {\rm R}_2 ~~~~~~~ ; \\
{\rm L}_3 & = & i \s^1 & \otimes & \s^1 & \otimes & \s^2 & \otimes & {\rm I} &
\otimes
& {\rm I} & \otimes & {\rm I} & = & {\rm R}_3 ~~~~~~~ ; \\
{\rm L}_4&=&i \s^1 & \otimes &\s^3& \otimes&\s^2 &\otimes &{\rm I} & \otimes &
{\rm I} & \otimes & {\rm I} & = & {\rm R}_4 ~~~~~~~ ; \\
{\rm L}_5 &=&i \s^3& \otimes &{\rm I}& \otimes&\s^2 &\otimes &\s^1 & \otimes &
{\rm I} & \otimes & {\rm I} & = & {\rm R}_5 ~~~~~~~ ; \\
{\rm L}_6&=&i \s^3 & \otimes &{\rm I}& \otimes&\s^2 &\otimes &\s^3 & \otimes &
{\rm I} & \otimes & {\rm I} & = & {\rm R}_6  ~~~~~~~ ; \\
{\rm L}_7&=&i \s^3 & \otimes &{\rm I}& \otimes &{\rm I} &\otimes &\s^2 &
\otimes &
\s^1 & \otimes & {\rm I} & = & {\rm R}_7 ~~~~~~~\, ; \\
{\rm L}_8&=&i \s^3 & \otimes &{\rm I}& \otimes&{\rm I}& \otimes &\s^2 & \otimes
&
\s^3 & \otimes & {\rm I} & = & {\rm R}_8 ~~~~~~~ ; \\
{\rm L}_9&=&i \s^3 & \otimes &{\rm I}& \otimes&{\rm I}& \otimes &\s^2 & \otimes
&
\s^2 & \otimes & \s^2 & = & {\rm R}_9  ~~~~~~~ ; \\
{\rm L}_{10}&=&i \s^3 & \otimes &{\rm I}& \otimes&\s^2 & \otimes &\s^2 &
\otimes &
\s^2 & \otimes & \s^1 & = & {\rm R}_{10} \, ~~~~~ ; \\
{\rm L}_{11}&=&i \s^3 & \otimes &{\rm I}& \otimes&\s^2& \otimes &\s^2 & \otimes
&
\s^2 & \otimes & \s^3 & = & {\rm R}_{11} \,  ~~~~~ ; \\
{\rm L}_{12}&=& {\rm I} & \otimes &{\rm I}& \otimes&{\rm I}& \otimes &{\rm I} &
 \otimes & {\rm I} & \otimes & {\rm I} & = & - \, {\rm R}_{12} ~~~  ;
\end{array} \eqno(A.6) $$
The case of N = 11 is contained as a subset.

At this point we should mention that real Dirac gamma matrices have been
discussed in a number of references. However, we  found a previous work by
De Crumbrugghe and Rittenberg \cite{DeCRT} very useful in our study and
construction of real generalized Pauli matrix representations.
The interested reader is referred to this prior work for additional
information.

\noindent The final explicit result that we present is for N = 16, where
we have a 128 x 128 representation of the N = 16 supersymmetry algebra:
$$\begin{array}{ccccccccccccccccrc}
{\rm L}_1 & =& i \s^2 &\otimes & {\rm I}& \otimes & {\rm I} & \otimes
& {\rm I} & \otimes & {\rm I} & \otimes & {\rm I} & \otimes & {\rm I} &
=&  {\rm R}_1 &; \\
{\rm L}_2 & =& i \s^{1} &\otimes &\s^1 &\otimes& \s^2 &\otimes &{\rm I}&
\otimes& {\rm I} & \otimes & {\rm I} & \otimes &{\rm I} &=&   {\rm R}_2&; \\
{\rm L}_3 & =& i \s^1&\otimes&{\rm I} & \otimes& \s^1& \otimes &\s^2 &\otimes
& {\rm I} & \otimes &{\rm I}& \otimes &{\rm I} &=&   {\rm R}_3 &; \\
{\rm L}_4 &=&i \s^1 &\otimes &{\rm I}& \otimes &\s^3 &\otimes &\s^2 &\otimes &
{\rm I}& \otimes & {\rm I} & \otimes & {\rm I} &=& {\rm R}_4& ;\\
{\rm L}_5 &=& i \s^1 &\otimes &\s^2 &\otimes& {\rm I}& \otimes &\s^1 &\otimes
&{\rm I}&\otimes & {\rm I}&\otimes &{\rm I}&=& {\rm R}_5& ; \\
{\rm L}_6 &=&i \s^1 &\otimes &\s^2 &\otimes& {\rm I}& \otimes &\s^3 &\otimes
&{\rm I}&\otimes& {\rm I} & \otimes &{\rm I}&=& {\rm R}_6&; \\
{\rm L}_7&=&i\s^1&\otimes&\s^2 &\otimes &\s^2 &\otimes &\s^2 &\otimes & {\rm I}
&\otimes& {\rm I}&\otimes &{\rm I} &=&{\rm R}_7&; \\
{\rm L}_8 &=&i \s^1& \otimes& \s^3& \otimes &\s^2 &\otimes&{\rm I}&\otimes
&{\rm I} &\otimes&{\rm I}&\otimes&{\rm I}&=&{\rm R}_8&;\\
{\rm L}_9 &=&i \s^3 &\otimes&{\rm I}&\otimes&{\rm I}&\otimes
&{\rm I} &\otimes &\s^2& \otimes &\s^2& \otimes &\s^2 &=&{\rm R}_9&;\\
{\rm L}_{10} &=& i \s^3 &\otimes& {\rm I}&\otimes& {\rm I}&\otimes
&{\rm I}& \otimes &\s^2 &\otimes& {\rm I}&\otimes &\s^3 &=&{\rm R}_{10}&;\\
{\rm L}_{11} &=& i \s^3 &\otimes& {\rm I}& \otimes& {\rm I}&\otimes
&{\rm I}&\otimes &\s^2 &\otimes&{\rm I}& \otimes &\s^1 &=& {\rm R}_{11}&;\\
{\rm L}_{12} &=&i \s^3 &\otimes&{\rm I}&\otimes&{\rm I}& \otimes
&{\rm I}&\otimes &\s^1 &\otimes &\s^2 &\otimes&{\rm I}&=&{\rm R}_{12}&;\\
{\rm L}_{13} &=& i \s^3& \otimes & {\rm I}& \otimes& {\rm I}& \otimes
&{\rm I}&\otimes&{\rm I}&\otimes&\s^1&\otimes &\s^2&=&{\rm R}_{13}&;\\
{\rm L}_{14} &=&i \s^3 &\otimes& {\rm I}& \otimes &{\rm I}&\otimes
&{\rm I}& \otimes &\s^3 &\otimes& \s^2 &\otimes& {\rm I}&=&{\rm R}_{14}&;\\
{\rm L}_{15} &=& i \s^3 &\otimes&{\rm I}&\otimes&{\rm I}&\otimes
&{\rm I}&\otimes&{\rm I}&\otimes&\s^3&\otimes&\s^2&=&{\rm R}_{15}&; \\
{\rm L}_{16} &=& {\rm I} & \otimes & {\rm I} & \otimes & {\rm I} & \otimes
& {\rm I}&\otimes& {\rm I}&\otimes &{\rm I} &\otimes
&{\rm I} &=& - {\rm R}_{16} & .
\end{array}
\eqno(A.7)$$
Similarly, N = 13, 14 and 15 can be formed using the matrices for the N = 16
case.

\sect{APPENDIX B: A Universal Theory of On-Shell D $\le$ 3 Scalar Multiplets}

One of the most interesting implications of our 1D results is that they
provide a general way to describe {\underline {all}} supersymmetric scalar
multiplets for D $\le$ 3! This includes the cases of
scalar multiplets in D = 3, D = 2 (both heterotic and non-heterotic) and D
= 1.  We now as well have, from the application of our work on
fermionic multiplets, a recipe for describing heterotic fermion
multiplets for arbitrary N.  Below for the scalar multiplets, we
explicitly consider the case in 3D (all other cases can be obtained
via truncations).

Let ${\cal A}_{i}(x)$ and $ \psi_{\a \, \hat j} (x)$ correspond to a multiplet
of real spin-0 and 3D, Majorana spinors, respectively.  We propose a
set of on-shell supersymmetry variations given by,
$$
\d_{Q} \, {\cal A}_{i}~=~ \, \e^{\a \, I} \, \left( {\rm L}_I \right)_i
{}^{\hat j} \,  \psi_{\a \, \hat j} ~~~, ~~~
\d_{Q} \, \psi_{\a \, \hat i}~=~i \, \e^{\b \, I} \, \left( {\rm R}_I
\right)_{\hat i} {}^{j} \, (\g^a)_{\a \b} \pa_{a} {\cal A}_{j} ~~~ ,
\eqno(B.1) $$
in terms of a constant 3D Majorana spinor parameter $\e^{\a \, K}$.
(Note that our convention for complex conjugation of spinors is:
${\left( \e^{\a} \right)}^*~=~ + \e^{\a}$~~ and ~~${\left( \e_{\a} \right)}^*~
{}~=~ -\e_{\a}$.)
A simple calculation reveals,
$$ \left[ \d_{Q} \left( \e_{1} \right), \d_{Q} \left( \e_{2} \right)
\right] \, {\cal A}_{i} ~=~ {\widehat \xi}_{1\, 2}^a \pa_a {\cal A}_i  ~~~,
\eqno(B.2)  $$
where ${\widehat \xi}_{1\, 2}^a \equiv i 2 \e^{\a \, I}_1 \e^{\b \, I}_2
(\g^a)_{\a \b}$.
Similarly, we have for the spinor field
$$ \left[ \d_{Q} \left( \e_{1} \right), \d_{Q} \left( \e_{2} \right)
\right] \,  \psi_{\a \, \hat i} ~=~  {\widehat \xi}_{1\, 2}^a \pa_a
\psi_{\a \, \hat i} ~+~ i \fracm 12   \e_{\a}{}^{ \, I}_{ [1 }
\e^{\b  \, J}_{ 2 ] } \left( {\rm R}_{J} {\rm L}_{I}
\right)_{\hat i} {}^{\hat j} \,  (\g^a)_{\b \g} \pa_{a} \psi^{\g}{}_{ \,
\hat j} ~~~.  \eqno(B.3)  $$
The latter term on the RHS of (B.3) is proportional to the equation
of motion for the spinor field and in an on-shell description we
set it equal to zero as is customary. The invariant on-shell action
takes the expected form,
$$
{\cal S}_{3D, \, SM} = \int d^3 x \, \fracm 12 [~ {\eta}^{a b} (\, \pa_a
 {\cal A}_i  \,) (\, \pa_b  {\cal A}_i \,) ~+~ i (\g^a)^{\b \g}
\psi_{\b \, \hat j} \pa_a \psi_{\g \, \hat j} ~]  ~~~. \eqno(B.4)
$$

We should mention that for 3D theories, supersymmetric scalar multiplets
are dual to supersymmetric vector multiplets. So our work as well provides
a basis for construction of 3D, arbitrary N-extended supersymmetric vector
multiplets described by fields (${\cal B}_i {}^j$, $\l_{\a \, \hat k}
{}^i$, $A_a$). These have variations given by,
$$\eqalign{
\d_Q {\cal B}_i {}^j &=~  \e^{\a \, I} \, ({\rm L}_I)_k {}^{\hat k} \,
\left[ ~ \d_i {}^k \l_{\a \, \hat k} {}^j ~-~ {\rm d}^{-1} \d_i {}^j
\l_{\a \, \hat k} {}^k   ~ \right] ~~~, \cr
\d_Q \l_{\a \, \hat k} {}^k &=~ i \e^{\b \, I} \, ({\rm R}_I )_{\hat k} {}^j
(\g^a )_{\a \b} ~\left[ \, \pa_a {\cal B}_j {}^k ~+~ \fracm 12 \, {\rm d}^{-1}
\,
\d_j {}^k \e_a {}^{b c} F_{b c}  \, \right] ~~~, \cr
\d_Q A_a &=~ i \e^{\a \, I} \, ({\rm L}_I)_k {}^{\hat k}  \, (\g_a )_{\a \b}
\, \l^{\b}{}_{  \hat k} {}^k   ~~~. } \eqno(B.5)$$
The spectrum as well as the supersymmetry variations are completely
determined by tensoring the $i$ representation with (B.1) and then
performing a duality transformation only on the resultant singlet
spin-0 field.  It is also possible to form a twisted version of both
the N-extended scalar and vector multiplets.

Thus, we see that the equations in (3.3) together with (3.8) provide a
{\underline {universal}} formalism for describing on-shell scalar multiplet
theories in D $ \le $ 3 for {\underline {all}} values of N (the degree
of extended supersymmetry).  We believe that this work is also of great
importance for future studies. The search for the off-shell versions
of these theories is clearly a topic that warrants study. Such work
may ultimately shed light on the off-shell problem for even higher
dimensional theories. Once again, we cannot help but wonder about
the nature of the N $\to \infty$ limit of the 3D theories. According to
at least one previous work on supersymmetric quantum mechanical
systems \cite{DeCRT}, N-extended supersymmetry imposes a constraint
such that for N $>$ 4 the Hamiltonian
is integrable!

\newpage

\end{document}